\documentclass[aps, prb, twocolumn]{revtex4}
\usepackage{amsmath}
\usepackage{graphicx}
\usepackage{epsfig}

\usepackage{bm}

\newcommand{\be}{\begin{equation}}
\newcommand{\ee}{\end{equation}}

\newcommand{\beq}{\begin{eqnarray}}
\newcommand{\eeq}{\end{eqnarray}}

\def\H1{\widehat{H}_1}

\begin{document}

\title{Superconductivity in the repulsive Hubbard model}
\author{D.\ Baeriswyl}
\affiliation{Dept.\ of Physics, University of
Fribourg, CH-1700 Fribourg, Switzerland}

\date{\today}

\begin{abstract}
The two-dimensional repulsive Hubbard model has been investigated by a variety of
methods, from small to large $U$. Superconductivity with $d$-wave symmetry is consistently found close to half filling. After a brief review of the various methods a variational many-electron state is discussed in more detail. This trial state is a natural extension of the Gutzwiller ansatz and provides a substantial improvement thereof.
\end{abstract}
\maketitle
\section{Introduction}
During the last two decades the Hubbard Hamiltonian has become {\it the}  model for strongly correlated particles, both fermions and bosons, and nowadays its investigation does not need to be justified. A great variety of phenomena have been interpreted in terms of the Hubbard model, such as ferromagnetism, antiferromagnetism, bond alternation, spin liquids, superconductivity and even charge stripes. Due to the complexity of the theoretical problem the analysis is typically based either on approximate treatments -- mean-field theory, perturbative expansions, variational wave functions -- or on ``numerically exact'' methods -- diagonalization of the Hamiltonian for small system sizes, quantum Monte Carlo, Density Matrix Renormalization Group. Each method has its own advantages, but also its drawbacks. Clearly one has to examine the various results with a critical mind because often the approximations are too crude and may thus overestimate certain ordering tendencies. A famous example is the Stoner theory of ferromagnetism. If applied to the simple Hubbard model on the square lattice at half filling, this theory predicts a ferromagnetic ground state for any (positive) value of the Hubbard parameter $U$. In contrast, detailed variational calculations show that there is no ferromagnetism at half filling (where the ground state is antiferromagnetic) and close to half filling ferromagnetic ordering could at most occur for extremely large values of $U$.  

While ferromagnetic long-range order fades away if methods are used that are more accurate than Stoner theory, the situation is radically different in the context of superconductivity. In fact, within BCS theory the repulsive Hubbard model is not superconducting. The on-site repulsion clearly is detrimental to on-site pairing, which can be discarded right away. For other types of pairing, such as extended $s$-wave, $p$-wave or $d$-wave, the on-site repulsion gives zero contribution to the condensation energy in BCS mean-field theory. Thus, if the Hubbard model has a superconducting ground state, then the order parameter has to be stabilized by electron correlations. A pictorial view of such an unusual superconductor has been given by Anderson in his theory of resonating valence bonds  where singlet bonds of the parent antiferromagnetic insulator turn into charged superconducting pairs upon doping \cite{Ande1}. Detailed recent studies for small, intermediate and large values of $U$ confirm that the repulsive Hubbard model on a square lattice does exhibit superconductivity with $d$-wave symmetry for electron densities close to (but not equal to) 1.

\section{From small to large $U$}
We consider the Hubbard Hamiltonian
\begin{equation}
\hat{H}=\hat{H}_0+U\hat{D}\, ,
\end{equation}
where
\begin{equation}
\hat{H}_0=-\sum_{ij\sigma}t_{ij}c_{i\sigma}^\dag c_{j\sigma}
\end{equation}
describes hopping over the sites of a square lattice, with the usual fermion creation and 
annihilation operators, $c_{i\sigma}^\dag$ and $c_{i\sigma}$, respectively, and
\begin{equation}
\hat{D}=\sum_{i}n_{i\uparrow}n_{i\downarrow},\quad \mbox{with} \quad
n_{i\sigma}=c_{i\sigma}^\dag c_{i\sigma}\, ,
\end{equation} 
measures the number of doubly occupied sites. Only hopping
between nearest ($t_{ij}=t$) and next-nearest neighbours ($t_{ij}=t'$) will be
considered. The bare spectrum,
\begin{equation}
\varepsilon_{\bf k}=-2t(\cos k_x+\cos k_y)-4t'\cos k_x\cos k_y\, ,
\label{tb}
\end{equation}
has a bandwidth $W=8t$. We discuss the three regimes of small $U$ ($U\ll W$),
intermediate $U$ ($U\approx W$) and large $U$ ($U\gg W$) separately because they are usually
approached with rather different techniques.

\subsection{Small $U$}
The question of Cooper pairing in a fermionic system with purely repulsive interactions has been addressed a long time ago by Kohn and Luttinger \cite{Kohn1} using a perturbative calculation of the effective two-particle vertex in powers of the coupling strength. Their estimate for the critical temperature in the case of $^3$He was deceptively small, but it was argued that in electronic systems with an appropriate band structure the situation may be more favorable \cite{Kohn1}. In the case of the Hubbard model on the square lattice close to half filling, divergences preclude the application of naive perturbation theory, but the development of the renormalization-group approach for interacting fermions during the early nineties paved the way for studying the effective two-particle vertex also in this case. Several groups used this technique, which consists in tracing out high-energy states in incremental steps and at the same time incorporating their effects within a renormalized vertex, which is a function of momenta \cite{Zanc1, Halb1, Hone1, Binz1}. These studies differ in technical details, but not in the main message, namely that at and very close to half filling the ground state of the two-dimensional Hubbard model is antiferromagnetic while for higher but not too high doping $d$-wave superconductivity prevails. It is not easy to provide accurate quantitative predictions for gap sizes, critical temperatures or order parameters using this approach, but for $U$ of the order of $t/2$ these quantities are not ridiculously small.

\subsection{The large $U$ limit}
In the large $U$ limit, where double occupancy is suppressed, the Hubbard model can be
replaced by the $t$-$J$ model close to half filling
\cite{Chao1, Faze1}, defined by the Hamiltonian
\begin{equation}
\hat{H}_{\mbox{\small t-J}}=-\sum_{ij\sigma}t_{ij}c_{i\sigma}^\dag c_{j\sigma}
+\sum_{ij}J_{ij}{\bf S}_i\cdot{\bf S}_j\, ,
\label{tJ}
\end{equation}
where ${\bf S}_i$ are spin $\frac{1}{2}$ operators and $J_{ij}=2t_{ij}^2/U$. This Hamiltonian is
supposed to act in the reduced space of configurations without doubly occupied sites. It is important to mention that in general additional three-site terms exist which are expected to be negligible only very close to half filling. Further away from half filling these terms should be kept if one wants to study the large $U$ limit of the Hubbard model. A great variety of techniques have been applied to the $t$-$J$ model 
\cite{Lee1, Ogat1}. Here we just mention a simple variational ansatz for the ground state,
the ``resonating valence bond state'' (RVB) of Anderson \cite{Ande1},
\begin{equation}
\vert\Psi_{\mbox{\small RVB}}\rangle=\prod_i(1-n_{i\uparrow}n_{i\downarrow})\vert\Psi_0\rangle\, ,
\label{rvb}
\end{equation} 
where $\vert\Psi_o\rangle$ represents the filled Fermi sea or a BCS singlet superconductor. 
Variational Monte Carlo results for $d$-wave pairing with a gap function
\begin{equation}
\Delta({\bf k})=\Delta_0(\cos k_x-\cos k_y)
\end{equation} 
yield a large region of doping, up to about 40\%, where this state is preferred \cite{Para1}. The gap parameter $\Delta_0$ is found to be largest at half filling, of the order of $t$,
at the same time the pair amplitude goes to zero when approaching half filling. This unusual
behavior has been associated with the pseudogap of underdoped cuprates \cite{Ande2}.

\subsection{Cuprates, a case for intermediate $U$}
The application of the Hubbard model to layered cuprates is {\it a priori} not obvious, but several
experiments indicate that this is a good starting point. Photoemission data exhibit a single band crossing the Fermi surface, at least in optimally and overdoped samples, and the shape of the
Fermi surface can be fitted rather well \cite{Dama1} using the tight-binding band structure (\ref{tb}) with $t'\approx -0.3t$. Inelastic neutron scattering for the parent antiferromagnetic insulator can be interpreted in terms of the Heisenberg model, provided that ring exchange is included
\cite{Cold1}. This term appears naturally in the Hubbard model if the large $U$ expansion is carried out to higher orders than in Eq.\ (\ref{tJ}). The comparison between theoretical and
experimental dispersion curves for magnetic excitations gives \cite{Kata1} an estimate of 
$U\approx W$.  The analysis of the neutron data is consistent with photemission experiments on the antiferromagnetic insulator, where the momentum distribution $n({\bf k})$ shows a pronounced 
${\bf k}$-dependence \cite{Dama1}, in contrast to the $t$-$J$ model where $n({\bf k})=1$.
We conclude that $U$ is of the order of the bandwidth $W$ where neither the functional renormalization group, valid for $U\ll W$, nor the $t$-$J$ model, representing the $U\gg W$ regime, can be trusted.

\subsection{Variational results for intermediate $U$}
Numerical approaches, such as quantum Monte Carlo \cite{Dago1} or quantum cluster methods   \cite{Maie1, Sene1} are widely used to unravel the secrets of the Hubbard model, especially for
intermediate values of $U$. These techniques have certain advantages, for instance to be rather unbiased, but they have also their problems, such as the trouble with the fermionic sign and/or the restriction to small cluster sizes. An alternative route is the use of variational wave functions. The RVB state (\ref{rvb}) is clearly not appropriate
for $U$ of the order of the bandwidth, where doubly occupied sites cannot be excluded. Many
other proposals have appeared in the literature, but we restrict ourselves to one of them,
defined as
\begin{equation} 
\label{gb}
\vert\Psi_{\rm GB}\rangle=\rm e^{-h\hat{H}_0/t}\,
\rm e^{-g\hat D}\vert\Psi_{0}\rangle\, .
\end{equation}
The operator $\rm e^{-g\hat D}$ partially suppresses double occupancy for $g>0$, 
while $\rm e^{-h\hat{H}_0/t}$ promotes both hole motion and kinetic exchange. The limit $h\rightarrow 0$ leads to the Gutzwiller ansatz. In the large $U$
limit and close to half filling the variational ansatz (\ref{gb}) for the Hubbard model is
equivalent to the RVB state (\ref{rvb}) applied to the $t$-$J$ model \cite{Baer1}, but we should keep in mind that the state (\ref{gb}) is a rather poor approximation for $U\gg W$ \cite{Dzie1, Baer1}.

The results obtained with the variational ground state (\ref{gb}) for $U=8t$, $t'=-0.3t$ can be summarized as follows \cite{Eich1}. 
Superconductivity occurs in a limited doping range, namely for electron densities
$0.75< n< 1$ on the hole-doped side and $1.05< n<1.2$ for electron doping. The two endpoints
0.75 and 1.2 correspond to the two specific densities beyond which the antiferromagnetic zone boundary does no longer cut the Fermi surface (the ``hot spots'' have disappeared). This strongly
hints at pairing induced by antiferromagnetic fluctuations. Very close to half filling the magnetic
correlations become so strong that they suppress both the superconducting gap parameter and the Cooper pair amplitude. There are
qualitative differences between hole and electron doping.  In the former case the energy gain comes clearly from the kinetic energy, while in the latter case the more conventional gain in potential energy is found. Moreover the condensation energy for the hole-doped side is much larger than that for electron doping. 

The variational results obtained with the ansatz (\ref{gb}) are in good 
agreement with quantum cluster calculations \cite{Kanc1} and experimental 
phase diagrams \cite{Kroc1}. A quantum Monte Carlo study by Aimi and 
Imada \cite{Aimi1}, often quoted as counter-argument against superconductivity 
in the repulsive Hubbard model, is also consistent with these variational
results, not however with the predictions of the RVB trial state (\ref{rvb}).

\section{Concluding remarks}
An important issue in the discussions of layered cuprates is the nature of 
the ``pseudogap phase''. In contrast to the RVB theory \cite{Para1, Ande2} 
the variational results discussed above did not show an energy scale steadily 
increasing when approaching half filling. This can be due to
the fact that competing instabilities, such as charge stripes, 
incommensurate spin-density waves or circulating currents have not been 
included in the ansatz. In any case, the superconducting gap, which was 
found to decrease when approaching half filling, is more naturally associated 
with an onset temperature for superconducting fluctuations than with the 
temperature $T^*$ below which the pseudogap appears.


\end{document}